\documentclass[aps,twocolumn,prb,superscriptaddress,amsmath,showpacs,tightenlines]{revtex4}
\usepackage{epsfig,graphicx,times}
\usepackage{amstext}
\usepackage{amsmath}
\usepackage{amssymb}
\usepackage{graphicx}
\usepackage{latexsym}
\usepackage{bm}

\begin{document}

\title{Fast generation of multiparticle entangled state for flux
qubits in a circle array of transmission line resonators with tunable coupling}

\author{Z.H. Peng}\email{zhihui_peng@riken.jp}
\affiliation{Advanced Science Institute, RIKEN, Wako, Saitama
351-0198, Japan}
\author{Yu-xi Liu}
\affiliation{ Institute of Microelectronics, Tsinghua University,
Beijing 100084, China} \affiliation{Tsinghua National Laboratory for
Information Science and Technology (TNList), Tsinghua University,
Beijing 100084, China}
\author{Y. Nakamura}
\affiliation{Advanced Science Institute, RIKEN, Wako, Saitama
351-0198, Japan}\affiliation{NEC Green Innovation Research
Laboratories, Tsukuba, Ibaraki 305-8501, Japan}
\author{J.S. Tsai}
\affiliation{Advanced Science Institute, RIKEN, Wako, Saitama
351-0198, Japan}\affiliation{NEC Green Innovation Research
Laboratories, Tsukuba, Ibaraki 305-8501, Japan}

\begin{abstract}
We study a one-step approach to the fast generation of
Greenberger-Horne-Zeilinger (GHZ) states in a circuit QED system
with superconducting flux qubits. The GHZ state can be generated in
about $10$ ns, which is much shorter than the coherence time of flux
qubits and comparable with the time of single-qubit operation.
In our proposal, a time-dependent microwave field is applied to
a superconducting transmission line resonator (TLR) and displaces
the resonator in a controlled manner,
thus inducing indirect qubit-qubit coupling without residual
entanglement between the qubits and the resonator. The design of a tunably
coupled TLR circle array provides us with the potential for extending this
one-step scheme to the case of many qubits coupled via several TLRs.
\end{abstract}
\pacs{03.67.Lx, 42.50.Dv, 03.67.Bg, 85.25.Cp}
\maketitle

\section{Introduction}
Entanglement lies at the heart of quantum mechanics and plays a key
role in quantum information processing. In quantum error correction,
quantum teleportation and quantum cryptography, the generation of
high fidelity multi-particle entangled states, such as the
Greenberger-Horne-Zeilinger (GHZ) state or two qubit Bell state, is
required~\cite{Nielsen}. Therefore, the preparation and verification
of the entangled states are of great practical importance in quantum
information processing systems.

Superconducting qubits~\cite{MSS,you-nature} are the most promising
candidate for realizing solid-state quantum information processing.
Generation of the GHZ state in superconducting quantum circuits is
consequently a highly important issue. While 14-particle and
10-particle entanglement have been experimentally demonstrated in
trapped-ion systems~\cite{Monz2011} and photonic
systems~\cite{Gao2010}, respectively, theoretical studies have
focused on generating the three-qubit GHZ state in superconducting
charge~\cite{wei2006}, flux~\cite{kim2008} and phase~\cite
{Galiautdinov2008} qubit circuits with direct qubit-qubit
interaction. Experimental demonstrations of entanglement have also
been limited to the cases of the
two~\cite{berkley2003,izamalkov2004,Niskanen2007,Plantenberg2007} or
three~\cite{xu2005,Neeley2010,DiCarlo2010,altomare2010,Fedorov2011}
particles so far. Therefore, how to generate multiqubit GHZ states
in superconducting quantum circuits is still an open question.

The circuit QED system~\cite{Girvin2008} provides a possibly
scalable method of realizing quantum information processing with
superconducting qubits. In such a system, the quantized microwave
field can act as a data bus to transfer information between qubits.
The generation of a multiparticle (three or more particles) GHZ
state using the system of the circuit QED has been proposed in
Refs.~\onlinecite{Zhu2005,Helmer2009,Bishop2009,Hutchison2009,Tsomokos2008,Wang2010}.
However, the generation of the GHZ state in
Refs.~\onlinecite{Helmer2009,Bishop2009,Hutchison2009} is based on
measurement, and is thus probabilistic. The probability of
generation of the GHZ state exponentially decreases with the number
of qubits. The proposal in Ref.~\onlinecite{Wang2010}, which is
similar to that for trapped-ion
systems~\cite{Molmer1999,Sorensen2000} and atomic
systems~\cite{Solano2003,Zheng2003}, comprises deterministic
generation of the GHZ state in a system of either flux qubits or
charge qubits. However, it is difficult to realize this approach in
current experiments due to the existence of a few practical
problems, as follows. First, to prepare a high-fidelity GHZ state,
the time control of the dc current pulse should be precisely set
around $2\pi/\omega_r$, where $\omega_r$ is the frequency of the
fundamental cavity mode and the usual range is from $2\pi\times 1$
GHz to $2\pi\times 10$ GHz. It is a major technical challenge to
carry out the proposed experiment with the present time resolution
of high-performance commercial arbitrary waveform generator, which
is generally about $1$ ns. Furthermore, changes in the half of the flux
quantum in an $\alpha$ loop (defined in Sec.~\ref{a_cavity}) with a
typical area, for example, $10\,\mu m^2$ also represent a technical
problem, because it is necessary to apply a biased current up to
$1\,$mA via on-chip biased line in an ultra-low temperature
environment if the mutual inductance between the $\alpha$ loop and
the biased line is $\sim1\,$pH. Second, the preparation time is one
or even two orders of magnitude longer than that of the single-qubit
operation. Faster preparation requires reduction of the frequency of
the fundamental cavity mode. This may lead to more operational
errors arising from the thermal excitations in the cavity. Third,
the number of superconducting flux qubits that can be placed around
current anti-node point of the one-dimensional superconducting
transmission line resonator (TLR) is limited, and thus the proposed
scalable method for many qubits remains an open question.

To overcome the problems encountered in previous studies (e.g., in
Ref.~\onlinecite{Wang2010}) and make the proposal experimentally
more feasible, we introduce a one-step multi-qubit GHZ state
generation method in the system of a circuit QED with flux qubits.
The advantages of our proposal are as follows. 1) The classical
driving field is directly applied to the TLR, in contrast to the
usual case in which the driving field is separately applied to the
qubits. Thus, the interactions between the qubits and the classical
field are induced by coherently displacing the cavity field. This
facilitates experiments not only to obtain homogenous coupling
constants, but also to realize synchronization of the  driving on
different qubits. 2) The generation time can be as short as the
single-qubit operation time. 3) Our proposal can be extended a case
of many qubits case by using a coupled TLR array as a data bus. We
expect that our proposal will work well for more than $20$ qubits
from our current experiments with the simplest setup, in which the
flux qubits are placed inside two tunably coupled TLRs via a
dc-SQUID.

Our paper is organized as follows. In Sec.~\ref{a_cavity}, one-step
generation of the GHZ state in flux qubits coupled to a TLR system
is presented using both analytical and numerical analysis. In
Sec.~\ref{coupledTLRs}, one-step generation of the GHZ state in flux
qubits coupled to two or more TLRs with tunable coupling is
discussed. Finally, discussion and conclusion are presented in
Sec.~\ref{Discussion}.

\section{A GHZ state generation for flux Qubits inside a cavity}\label{a_cavity}

\subsection{Theoretical Model}
\begin{figure}
\includegraphics*[scale=0.5]{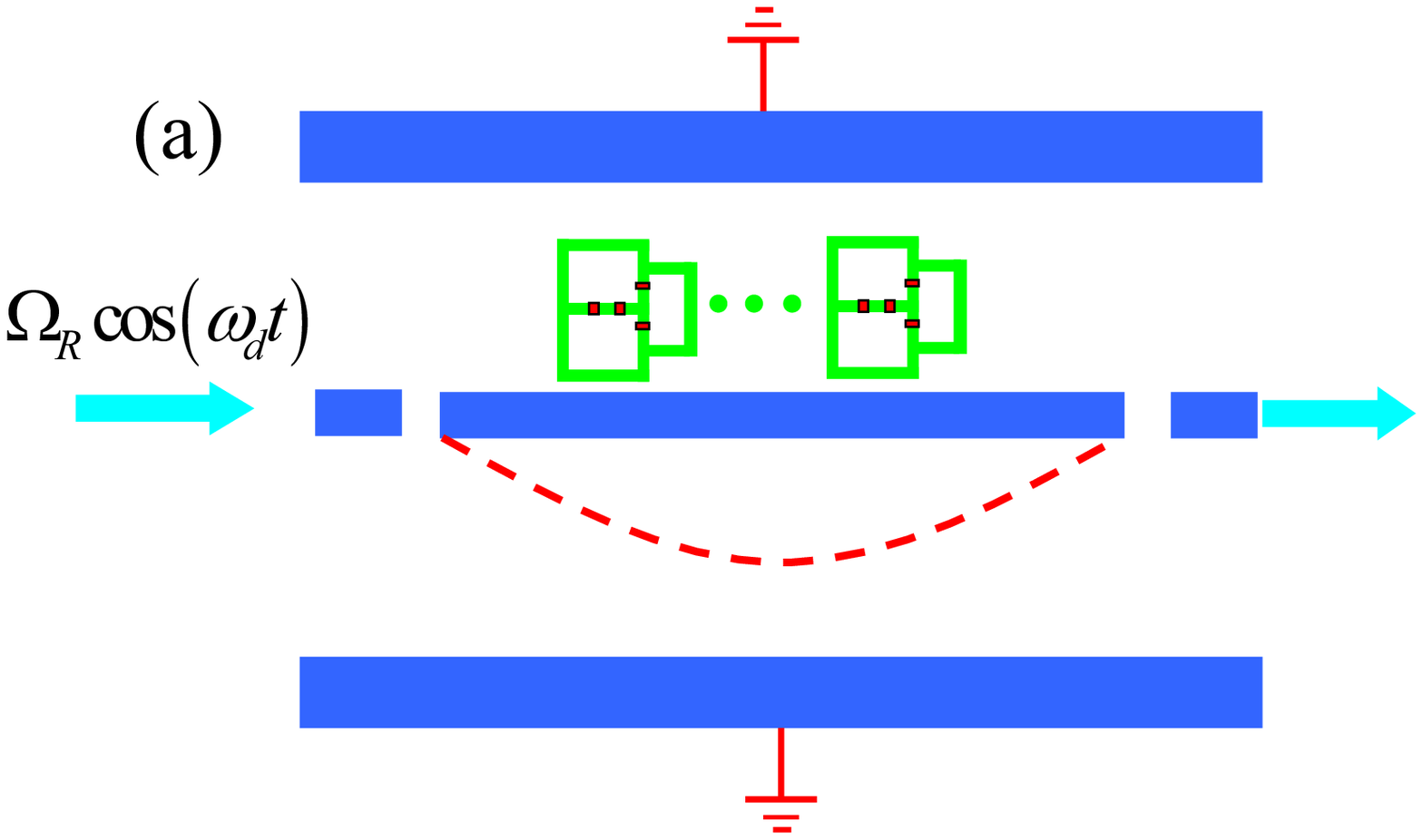}
\includegraphics*[scale=0.5]{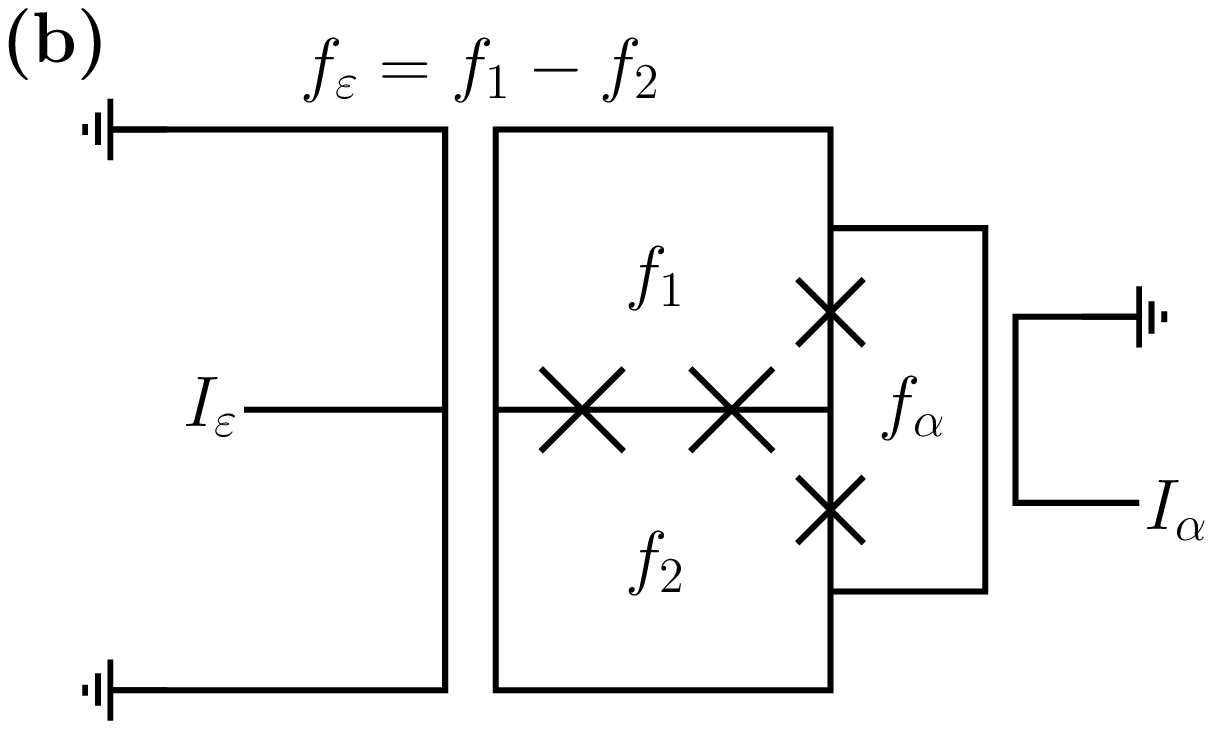}
\includegraphics*[scale=0.5]{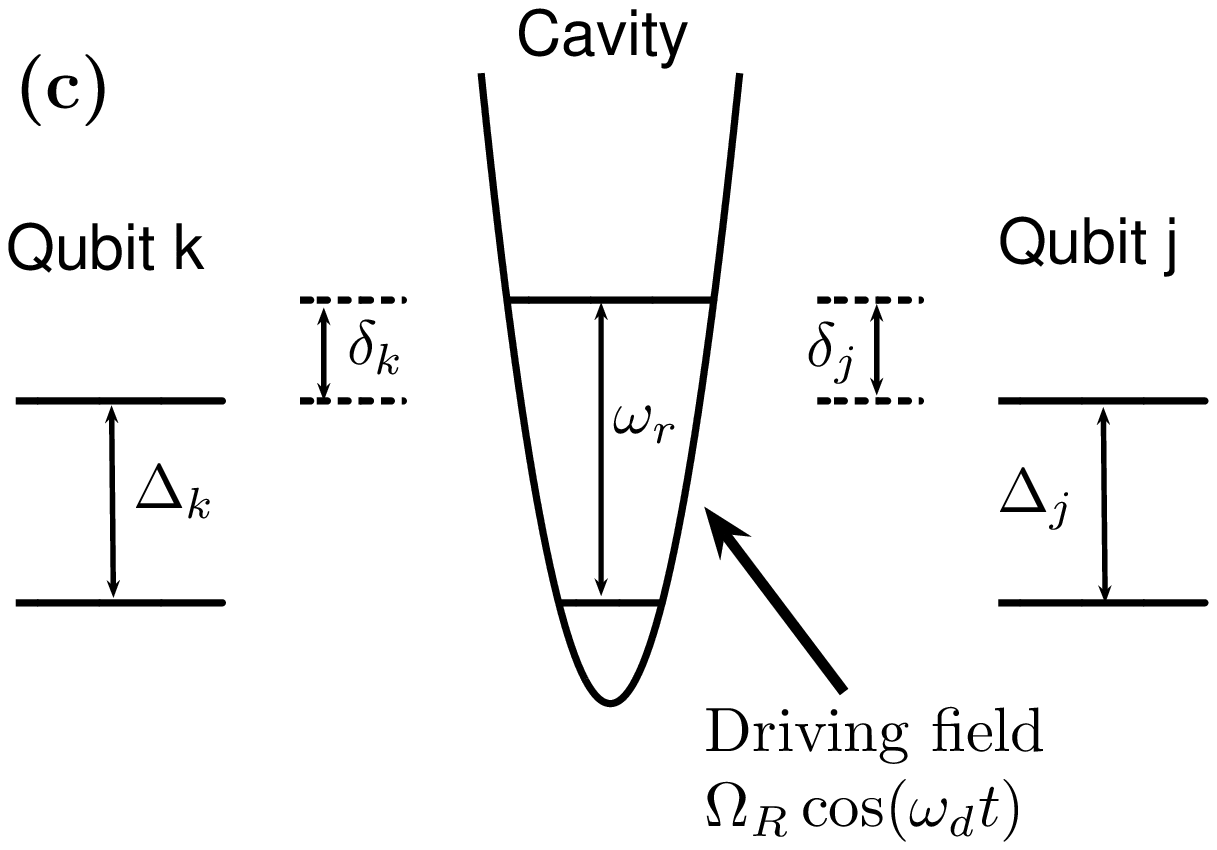}
\caption{Schematic description of our setup. (a) The flux
qubits are placed around current anti-node of the TLR (half
wavelength or full wavelength). The red dashed curve illustrates the
amplitude of the magnetic field of the half-wavelength TLR. (b) Schematic description of the tunable gap gradiometric flux qubits.
The crosses denote Josephson junctions. The qubit is controlled by the
bias lines $I_{\varepsilon}$ and $I_{\alpha}$. (c) Energy-level
diagram of the system. $\Delta, \omega_d, \omega_r$ are energy gap
of the qubits, driving microwave frequency, and fundamental frequency of
the cavity mode, respectively. The strong classical driving field
denoted by the large arrow resonantly interacts with the flux qubits. We assume that all qubits have the same Rabi frequency $\Omega_R$.
}\label{oneresonatorscheme}
\end{figure}

The investigated system is schematically shown in Fig.~\ref{oneresonatorscheme}, where superconducting flux qubits are strongly coupled to a one-dimensional transmission line resonator (TLR) with the geometric length $L_{0}$, distributed inductor $L$, and capacitance $C$. We consider the fundamental mode of the TLR, which can be modeled as a simple harmonic oscillator with the Hamiltonian
\begin{equation}\label{TLRHamiltonian}
H_{\rm TLR}=\omega_{r}a^{\dagger}a,
\end{equation}
where we have set $\hbar=1$, the frequency of the fundamental mode
is given by $\omega_{r}/2\pi=1/\sqrt{LC}$, $a^\dagger$ and $a$ are
the creation and annihilation operators of the fundamental mode of
the resonator.

As shown in Fig.~\ref{oneresonatorscheme}(b), we assume that each flux qubit has a gradiometric configuration as studied in Ref.~\onlinecite{Paauw2009}. The gradiometric design is used to trap an odd number of fluxoids to bias the flux qubit at its optimal point, and also to greatly reduce environmental noise. In contrast to the flux
qubit with three junctions, one of which has a critical current $\alpha$ times smaller than that of the two identical junctions, here the small $\alpha$-junction is replaced by a so-called $\alpha$-loop, formed by a SQUID with two identical Josephson junctions. In this case, the ratio $\alpha$ and thus the coupling strength $\Delta$ between two circulating current states can be tuned via the magnetic flux $f_{\alpha}\Phi_{0}$ through the SQUID loop, where $\Phi_0$ is the magnetic flux quantum. The flux qubit Hamiltonian can be written as
\begin{equation}
H_{qb}=-\frac{1}{2}[\varepsilon\bar{\sigma}_{z}+\Delta(f_{\alpha}\Phi_{0})\bar{\sigma}_x],
\end{equation}
where the Pauli matrices read $\bar{\sigma}_z=|0\rangle\langle0|-|1\rangle\langle1|$ and $\bar{\sigma}_x=0\rangle\langle1|+|1\rangle\langle0|$, in which $|0\rangle$ and $|1\rangle$ are the clockwise and  counterclockwise persistent current states. $\varepsilon(f_{\varepsilon},f_{\alpha})=2I_{p}f_{\varepsilon}\Phi_0$ is the biased magnetic energy where $I_p$ is the persistent current in the qubit and $f_{\varepsilon}=f_1-f_2$ is the magnetic frustration difference in the two loop halves of the gradiometer. The energy gap $\Delta(f_{\alpha}\Phi_{0})$ can be controlled through the external magnetic flux $f_{\alpha}\Phi_{0}$.

Let us now assume that the flux qubits are placed around the current anti-node of the TLR. The coupling between qubits and the TLR is approximately homogeneous because the dimension of the qubits is on the order of several micrometers, which is much smaller than the wavelength of a few centimeters for the fundamental electromagnetic
modes in microwave frequency regime. We also assume that the distance between the two nearest flux qubits is sufficiently large ($\sim 80 \,\mu m$) to ensure that there are no direct interaction between different qubits. The coupling strength between the $k$-th flux qubit and the TLR is given as $g_k=M_kI_{p}^{k}I_{r0}$, where $M_k$
is the mutual inductance between the qubit and the resonator, $I_{r0}=\sqrt{\hbar\omega_r/L}$ is the zero-point current in the resonator.

The flux qubits are assumed to be near the optimal point ($\varepsilon\approx0$), and thus the total system Hamiltonian is as follows (refer to see Appendix A for detailed derivations):
\begin{equation}\label{SystemHamiltonian}
H_s=H_0+H_I+H_d,
\end{equation}
with

\begin{eqnarray}\label{SystemHam}
    H_0=\omega_r
    a^{\dagger}a+\sum\limits_{k=1}^{N}\frac{\Delta_k}{2}\bar{\sigma}_{x}^{k},\label{freeHam} \\
    H_I=\sum\limits_{k=1}^{N}g_k(a^++a)\bar{\sigma}_{z}^{k}, \label{intHam} \\
    H_d=\sum\limits_{k=1}^{N}\Omega_R\cos(\omega_dt)\,\bar{\sigma}_{z}^{k}.
\end{eqnarray}
Here, $H_0$ is the free Hamiltonian of the qubits and the cavity mode, $H_I$ is the interaction Hamiltonian between the qubits and the cavity mode, and $H_d$ describes the interaction between the qubits and classical driving microwave field. We assume that all qubits have the same frequency $\omega_d$ of the driving microwave field
and the same Rabi frequency $\Omega_R$. It should be noted here that the homogenous coupling $\Omega_R$ is induced by the classical field that is applied to the TLR (see Appendix A).

In the basis of the eigenstates of the qubits and neglecting fast oscillating terms using the rotating-wave approximation\,(RWA) if $g_k\ll\omega_r$ and $\Omega_R\ll\omega_d$, the Hamiltonian in Eq.~(\ref{SystemHamiltonian}) takes the form
\begin{eqnarray}\label{driving}
\widetilde{H_1}&=&\omega_ra^+a+\sum\limits_{k=1}^{N}\frac{\Delta_k}{2}\sigma_{z}^{k}+\sum\limits_{k=1}^{N}g_k(a^+\sigma_{-}^{k}+a\sigma_{+}^{k})
\nonumber \\
&+&\sum\limits_{k=1}^{N}\frac{\Omega_{R}}{2}(\sigma_{+}^{k}e^{-i\omega_d
t} +\sigma_{-}^{k}e^{i\omega_d t}).
\end{eqnarray}

Because the energy gap $\Delta_{k}$ for each qubit can be tuned by the biased flux in the $\alpha$ loop,  without loss of generality, we assume the energy gap of flux qubits are equal, i.e., $\Delta_{k}\equiv\Delta$. Thus, in the rotating reference frame at the frequency $\omega_d=\Delta$, the Hamiltonian in Eq.~(\ref{driving}) is changed to
\begin{equation}\label{drivingrotating}
\widetilde{H_2}=\delta\,
a^+a+\sum\limits_{k=1}^{N}g_k(a^+\sigma_{-}^{k}+a\sigma_{+}^{k})+\sum\limits_{k=1}^{N}\frac{\Omega_{R}}{2}(\sigma_{+}^{k}
+\sigma_{-}^{k}),
\end{equation}
where we have defined the detuning $\delta=\omega_r-\omega_d >0$ between the cavity field and the driving field, and the ladder operators $\sigma_{+}=|-\rangle\langle+|$ and $\sigma_{-}=|+\rangle\langle-|$ by using ground $|+\rangle$ and first exited $|-\rangle$ states of qubits. (If $\delta<0$, it only means the transition frequency of flux qubits is larger than the frequency of the fundamental cavity mode and also works well in our scheme.) Here, we assume that the circuit QED system works in small detuning regime, i.e., $\delta>g_k$ (not $\delta\gg g_k$). The third term of the Hamiltonian in Eq.~(\ref{drivingrotating}) implies the free Hamiltonian of the dressed qubits by the driving field~\cite{liu-2006}.

In the interaction picture with the unitary transformation
\begin{equation}
U(t)=\exp(-iH_{\rm fr}t)
\end{equation}
for the free Hamiltonian
\begin{equation*}
    H_{\rm fr}=\delta
    a^+a+\sum\limits_{k=1}^{N}\frac{\Omega_{R}}{2}(\sigma_{+}^{k}+\sigma_{-}^{k}),
\end{equation*}
the interaction part of the Hamiltonian in
Eq.~(\ref{drivingrotating})
\begin{equation*}
    H_{\rm int}=\sum\limits_{k=1}^{N}g_k(a^+\sigma_{-}^{k}+a\sigma_{+}^{k}).
\end{equation*}
becomes
\begin{eqnarray}\label{eq:10}
  \widetilde{H_{3}} & = & U^{\dagger}(t)H_{\rm int}U(t)\nonumber \\
  & = & \sum\limits_{k=1}^{N}\frac{g_k}{2}\{e^{-i\delta t}a[\sigma_x^k+\frac{1}{2}(\sigma_z^k-i\sigma_{y}^{k})e^{i\Omega_{R} t}
  \nonumber \\
  &   &-\frac{1}{2}(\sigma_z^k+i\sigma_{y}^{k})e^{-i\Omega_{R} t}]\}+H.c.  \label{dropHamiltonian}
\end{eqnarray}

In the strong-driving regime, i.e., $\Omega_R\gg \delta, \, g_k $, we can neglect fast oscillating terms and the Hamiltonian in Eq.~(\ref{eq:10}) turns into
\begin{small}
\begin{eqnarray}\label{effective Hamiltonian}
  H_{\rm eff}&=&
  \sum\limits_{k=1}^{N}\frac{g_k}{2}\sigma_x^k\{ae^{-i\delta t}+a^{\dagger}e^{i\delta
  t}\}.
\end{eqnarray}
\end{small}
The time evolution operator for the Hamiltonian in Eq.~(\ref{effective Hamiltonian}) can be written~\cite{Wei1963,wang2001} as
\begin{eqnarray}
U(t)&=& \prod\limits_{k\neq
j}^{N}\exp\left\{-\int_0^tB_k^*(t)dB_k(t)\sigma_x^k\sigma_x^j
\right\}
\\
    & & \prod\limits_{k}^{N}\exp\{-iB_k^*(t)a \sigma_x^k\}\prod\limits_{k}^{N}\exp\{-iB_k(t)a^+
    \sigma_x^k\}, \nonumber
\end{eqnarray}
with
\begin{equation}
    B_k(t)=\frac{ig_k}{2\delta}(e^{i\delta t}-1).
\end{equation}

It is obvious that $B_{k}(t)$ is a periodic function and vanishes at $t=T_n=2n\pi/\delta$ with the integer $n$. At these times, $U(t)$ is independent of the variables of the cavity field and the flux qubits are decoupled from cavity field. We define
\begin{equation}\label{gamma}
    \gamma_{kj}(t)\equiv \frac{1}{i}\int_0^tB_k^*(t)dB_j(t)
    =\frac{g_k g_j}{4\delta}\left[t-\frac{1}{i\delta}\left(e^{i\delta t}-1\right)\right].
\end{equation}

At time $T_n$, the time evolution operator takes the
form
\begin{eqnarray}
 U(T_n,\gamma_{kj})&=&\exp\left(-i\sum\limits_{k\neq
 j}^{N}\gamma_{kj}(T_n)\sigma_x^k\sigma_x^j\right)\nonumber\\
 &=&\exp\left(-i\sum\limits_{k\neq j}^{N}\frac{n\pi
 g_kg_j}{2\delta^2}\sigma_x^k\sigma_x^j\right).
\end{eqnarray}

For the convenience of the discussions, let us now assume that the
qubits are equally coupled to the resonator, i.e.,
$g_{k}=g_{j}\equiv g$, and thus we assume $\gamma_{kj} \equiv
\gamma$. It should be noticed that the inhomogenous coupling is not
a significant problem in our scheme. Actually, the homogenous
coupling is not really required for this type of gate. The extra
phase resulted from the inhomogenous coupling can be easily
corrected by single qubit operations as mentioned in
experiments~\cite{Leibfried2003}. If we adjust the parameters such
that $\gamma=(1+2m)\pi/8$ with an arbitrary integer $m$, then the
initially unentangled state
$|\psi(0)\rangle=\otimes_{k=1}^N|+\rangle_k$ can be changed to the
GHZ state with the unitary evolution $U(\gamma)$, here
$|\pm\rangle=(|0\rangle\pm|1\rangle)/\sqrt{2}$. The parameter
$\gamma$ is a geometric phase which will be addressed elsewhere. If
the parameters are selected as $n=1$, $m=0$, $g_k=2\pi\times50$ MHz,
$\delta=2 g_k$, then the GHZ state
\begin{equation}
|{\rm GHZ}\rangle=\frac{1}{\sqrt{2}}\left(\bigotimes\limits_{k=1}^N|
+\rangle_k+e^{i\pi(N+1)/2}\bigotimes\limits_{k=1}^N|-\rangle_k\right)
\end{equation}
is produced at the time $T=10$ ns. We notice that there are two theoretical works on realization of controlled phase gate based on the $\gamma$ with superconducting qubits~\cite{Yang2010,Wu2010}.

\subsection{Numerical Simulation} \label{Simulation1}
\begin{figure}
\includegraphics*[scale=0.4]{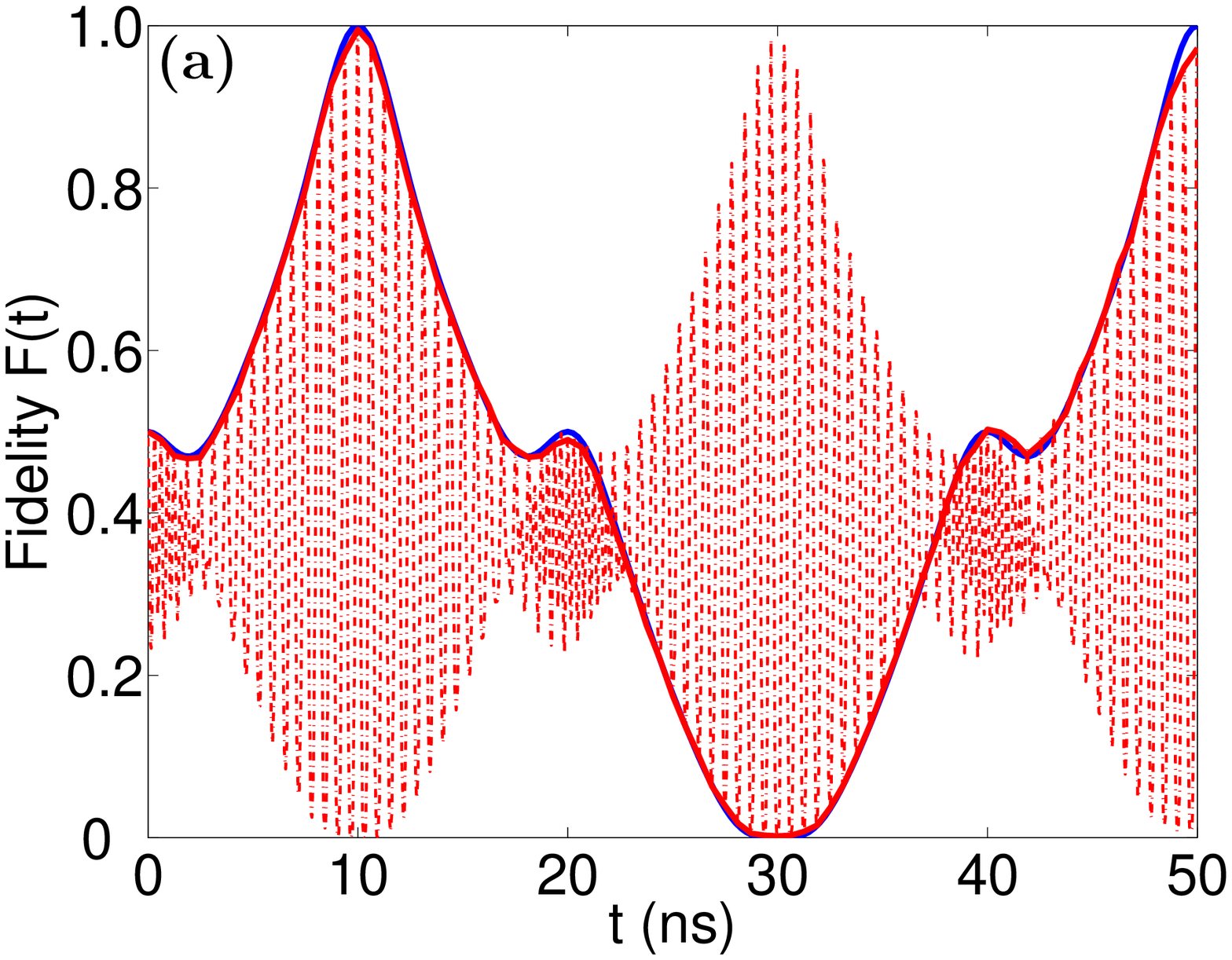}
\includegraphics*[scale=0.4]{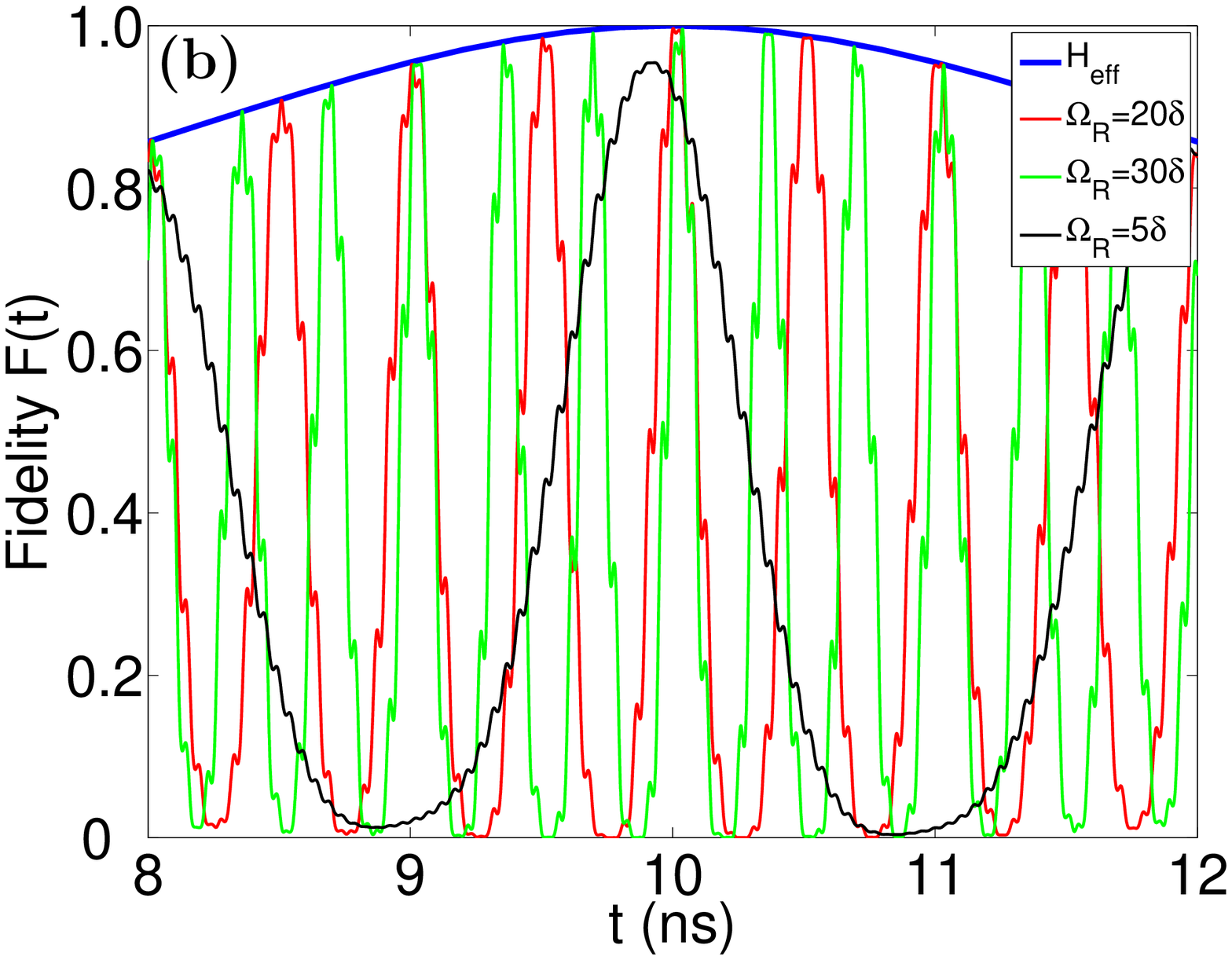}
\caption{Time dependence of the fidelity of the generated entangled
state in two flux qubits coupled to one TLR. The following parameters were used: $\omega_r=2\pi\times$10$\,$GHz,
$\Delta=2\pi\times10.1\,$GHz, $g=2\pi\times50\,$MHz,
$\delta=-2\pi\times100\,$MHz. The target EPR state is
$(|++\rangle+i|--\rangle)/\sqrt{2}$. (a) The blue solid, red solid,
and red dashed curves are simulated using the effective Hamiltonians
in Eq.~(\ref{effective Hamiltonian}) and in
 Eq.~(\ref{dropHamiltonian}) and the full Hamiltonian in
Eq.~(\ref{fullsimoneTLR}), respectively, with the optimized driving
strength $\Omega=20\delta$. (b) Simulation using the full
Hamiltonian in Eq.~(\ref{fullsimoneTLR}) for different driving
strengths around the time $t=2\pi/\delta=10\,$ns.
}\label{InitialHamiltonian}
\end{figure}

To verify the validity of the approach proposed here, we now present numerical calculations. We simulate the dynamics of the system by using both its full Hamiltonian in Eq.~(\ref{SystemHamiltonian}) without making any approximation, and the effective Hamiltonians in Eqs.~(\ref{dropHamiltonian}) and (\ref{effective Hamiltonian}). For the convenience, we can rewrite the full Hamiltonian in Eq.~(\ref{SystemHamiltonian}), in the basis of the eigenstates of the qubits and the rotating reference frame of $\omega_d$, as

\begin{eqnarray}\label{fullsimoneTLR}
\widetilde{H}_{\rm s} &=&
\widetilde{H_2}+\sum\limits_{k=1}^{N}\frac{\Omega_{R}}{2}(\sigma_{+}^{k}e^{2i\omega_d
t} +\sigma_{-}^{k}e^{-2i\omega_d
             t}) \\
&+&\sum\limits_{k=1}^{N}g_k(a^+\sigma_{+}^{k}e^{i(\omega_r+\omega_d)t}
+a\sigma_{-}^{k}e^{-i(\omega_r+\omega_d)t}).\nonumber
\end{eqnarray}

Let us assume that the cavity field is initially in the vacuum state $|0\rangle_{c}$, and the $k$-th qubit is initially in the state $|+\rangle_{k}=(|0\rangle+|1\rangle)/\sqrt{2}$, that is, the initial state of the whole system is
\begin{equation}
|\psi(0)\rangle=\bigotimes_{k=1}^N|+\rangle_k|0\rangle_{c}
=\bigotimes_{k=1}^N\frac{1}{\sqrt{2}}\left(|0\rangle_k+|1\rangle_k\right)|0\rangle_{c}.
\end{equation}

We can solve the Schr\"odinger equation to obtain the state $|\psi(t)\rangle$ at any time by using the Hamiltonians in Eq.~(\ref{dropHamiltonian}),  Eq.~(\ref{effective Hamiltonian}) and Eq.~(\ref{fullsimoneTLR}). Then we compare these states $|\psi(t)\rangle$ with the expected ideal GHZ state by using the fidelity
\begin{equation}
F(t)=Tr[\rho_{\rm GHZ}\rho_{q}(t)],
\end{equation}
where $\rho_{q}(t)$ is the reduced density matrix of the $N$ qubits and $\rho_{\rm GHZ}$ is the density matrix of the $N$-qubit GHZ state.

As an example, let us consider the interaction between two qubits and the cavity field with the initial state
$|++\rangle|0\rangle_{c}$. In Fig.~\ref{InitialHamiltonian}, the fidelities $F(t)$ of different outcomes versus the evolution time $t$ are plotted. In Fig.~\ref{InitialHamiltonian}(a), we find that both the full Hamiltonian
in Eq.~(\ref{fullsimoneTLR}) and the effective Hamiltonians in either Eq.~(\ref{dropHamiltonian}) or Eq.~(\ref{effective Hamiltonian}) well describe the dynamics of the two-qubits system. The expected entangled state $(|++\rangle+i|--\rangle)/\sqrt{2}$ can be generated in $ t=10$ ns, which is comparable with the single-qubit operation time.

To illustrate how the driving strength affects the fidelity of the expected GHZ state, we plot Fig.~\ref{InitialHamiltonian}(b) by using the Hamiltonian in Eq.~(\ref{fullsimoneTLR}) for different driving strengths. It can be clearly seen that the counter-rotating term of the driving field becomes important and reduces the fidelity when the driving strength is too strong. On the other hand, if the driving is too weak and the condition $\Omega_R\gg{g_k, \delta}$ is not met any more, the fast oscillating terms in Eq.~(\ref{dropHamiltonian}) must be taken into account. Therefore, the driving field has to be optimized so that the maximum fidelity can be achieved. We provide a more detailed analysis of the effect of these fast oscillating terms on the fidelity of the generated state below. With the optimized driving strength, the fidelity is at best above $99.5\%$. It should be noted here that we did not make an adiabatic approximation for the fast variable $\Omega_R$ in our simulation. Therefore, to prepare high fidelity GHZ states, the accuracy of control of the microwave pulse time should be determined by $2\pi/\Omega_R$ in our scheme from analysis of the full system Hamiltonian. For example, to realize the preparation with a fidelity exceeding $90\%$, the precision of the microwave pulse time should be around $100\,$ps for the detuning $\delta=-2\pi\times100\,$MHz and $\Omega_R=2\pi\times2\,$GHz in Fig.~\ref{InitialHamiltonian}(b). This is easily realized with a commercial pulse generator having a precision of $10\,$ps.

In Ref.~\onlinecite{Wang2010}, to increase the effective coupling between qubits, they have to increase $g$ (the coupling strength between qubits and resonator) and decrease the resonator frequency $\omega_r$. We argue that it is valid in experiments to make an adiabatic approximation for the resonator frequency $\omega_r$ when $g$ is approaching $\omega_r$ as in Fig.~2 of Ref.~\onlinecite{Wang2010}. Otherwise, to prepare high fidelity GHZ states in their scheme, the accuracy of control of the dc current pulse time is determined by $2\pi/\omega_r$, not by $2\pi\omega_r/g^2$. It means that the accuracy must be on the order of a few hundred ps for the selected parameters $\omega_r=2\pi\times1\,$GHz and $g=2\pi\times144\,$MHz in Fig.~2 of Ref.~\onlinecite{Wang2010}. Such accuracy would present a major challenge for state-of-the-art dc current pulse technology.

\subsection{Nonideal Case}

In the derivation of effective Hamiltonian in
Eq.~(\ref{dropHamiltonian}), we have neglected the following terms
\begin{eqnarray}\label{Neglectterm}
H_n &=&
\sum\limits_{k=1}^{N}\frac{g_k}{4}\left\{\left[(\sigma_z^k-i\sigma_{y}^{k})e^{i\Omega_{R}
t}\right.\right.\nonumber \\
&-&\left.\left.(\sigma_z^k+i\sigma_{y}^{k})e^{-i\Omega_{R}
          t}\right]ae^{-i\delta t}+H.c.\right\}   \\
          &=&\sum\limits_{k=1}^{N}\frac{ig_k}{2}\left[\sin(\Omega_R t)\sigma_z^k-\cos(\Omega_R
          t)\sigma_{y}^k\right]ae^{-i\delta t}+H.c.\,.\nonumber
\end{eqnarray}
These terms could reduce the fidelity of the gate. However, we can use the spin-echo technique to eliminate the errors from the $\sigma_z$ terms. Below, we will only study the effect of the $\sigma_y$ terms on the gate operation using the method described in Ref.~\onlinecite{Sorensen2000}.

In the interaction picture, the interaction Hamiltonian is $H_{n,I}(t)=U^{\dagger}(t)H_{n}(t)U(t)$, and we have the propagator $U_I(t)$ from the Dyson series,

\begin{eqnarray}
U_I(t) &=& 1-i\int_0^tdt'H_{n,I}(t')   \nonumber \\
       & &-\int_{0}^{t} \int_0^{t'}dt'dt''H_{n,I}(t')H_{n,I}(t'')+\cdots.
\end{eqnarray}
We can treat $U(t)$ as a constant during the integration because $H_n(t)$ is oscillating much faster than the propagator. Then, we get
\begin{eqnarray}
U_I(t) &=& 1-\frac{g}{2\Omega_R}\sin(\Omega_R
t)\sum\limits_{k=1}^{N}\{U^{\dagger}(t)\sigma_y^kU(t)\}  \nonumber \\
         & & +i\frac{g^2}{4\Omega_R^2}\sum\limits_{k\neq j}^{N}\{(1-\cos(2\Omega_R
         t))U^{\dagger}(t)\sigma_y^k\sigma_y^jU(t)\} \nonumber \\
         & & +\cdots.
\end{eqnarray}

Near the time $t=2\pi n/\delta$, $U(t)\approx\exp(-i\sum\limits_{k\neq j}^{N}\gamma\sigma_x^k\sigma_x^j)$ and we
obtain the fidelity
\begin{equation}\label{Fidelity}
   F(t)\approx1-\frac{N(N-1)g^2}{8\Omega_R^2}(1-\cos(2\Omega_R
         t)),
\end{equation}
where $N$ is the number of qubits involved in the gate. It should be noted that the estimation of $F(t)$ is obtained in the interaction picture. In the rotating frame and near the time $t=2\pi n/\delta$, the time evolution operator takes the form
\begin{equation}\label{timeoperator}
\widetilde{U}(t)\approx\exp(-i\sum\limits_{k}^N\Omega_R t\sigma_x^k/2)\exp(-i\sum\limits_{k\neq j}^{N}\gamma\sigma_x^k\sigma_x^j).
\end{equation}

In the experiment, we can accurately control the duration of classical microwave field to fulfill the condition both $\delta t=2n\pi$ and $\Omega_R t=4n'\pi$ ($n$ and $n^{\prime}$ are arbitrary integers) so that the effect of the $\sigma_y$ terms and $\exp(-i\sum\limits_{k}^N\Omega_R t\sigma_x^k/2)$ vanishes. When $N$ is not too large, e.g., $N<\Omega_r/g$, $F(t)$ is mainly limited by the accuracy of control of the microwave pulse. On the other hand, if the accuracy of control of the microwave pulse is fixed, there is a polynomial decrease of the fidelity according to $N$.

Our scheme works in the strong-driving regime, i.e., $\Omega_R\gg\delta,g_k$, so we can select the optimized driving strength $\Omega_R=20\delta=40g$ from the numerical simulation in Fig.~\ref{InitialHamiltonian}(b) and use Eq.~(\ref{Fidelity}) and Eq.~(\ref{timeoperator}) to roughly estimate the fidelity of the generated multi-particle GHZ state. In this case, $F(t)\sim 1$ near the time $t=2\pi/\delta=10\,$ns for $N<40$. We can conservatively expect
that our one-step proposal will work well for more than 20 qubits are placed at the current anti-node of the resonator by considering the inhomogenous coupling between qubits and resonator.

\section{Scalable circuit with tunable TLRs and tunable nodes}\label{coupledTLRs}
\subsection{Scalable model}
\begin{figure*}
\includegraphics[scale=0.8]{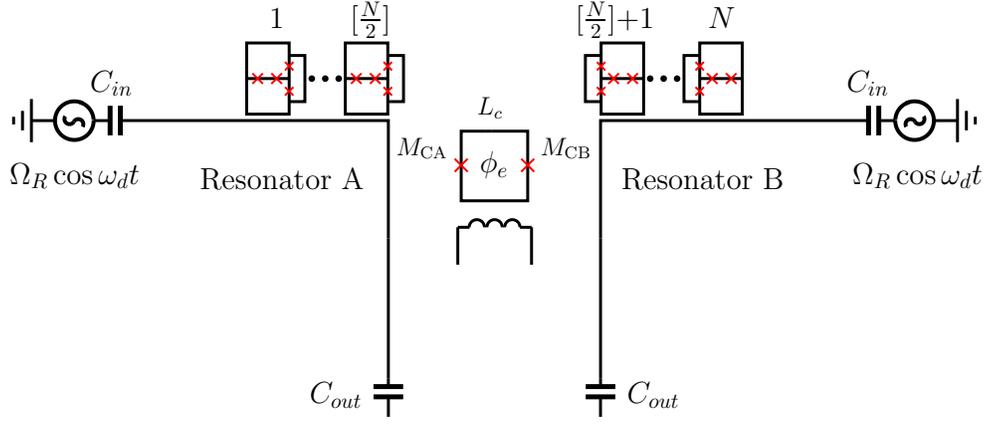}
\caption{Schematic description of flux qubits coupled
to two tunably coupled TLRs (half-wavelength TLRs or full-wavelength
TLRs). $N$ qubits are coupled to the current anti-node of the two
resonators. A dc-SQUID is placed at the current anti-node of the two
resonators. The inductive coupling between two resonators could be tuned by
changing the biased flux $\phi_e$ in the symmetric dc-SQUID loop. Two classical
microwave fields resonantly interact with $N$ qubits by driving the TLRs. $L_c$, $M_{\rm CA}$, $M_{\rm CB}$ are self-inductance of the dc-SQUID coupler loop, mutual inductance between the coupler and Resonator A, and mutual inductance between the coupler and Resonator B, respectively. }\label{twocoupledTLR}
\end{figure*}

In our design, the half wavelength of the TLR for the fundamental mode frequency $\omega_r=2\pi\times10$ GHz is around $6$ mm~\cite{Abdumalikov2008}. If the distance between the two nearest flux qubits is around $80$ $\mu$m, we can only place about $6$ flux qubits around the current anti-node position where the current variation is about $0.5\%$. Although the electric field is not completely zero around the anti-node, its effect on flux qubits is
negligibly small. It is possible to place more than 6 flux qubits in the half wavelength TLR because we can, in principle, design the coupling inductance so that the coupling constant $g_k$ is uniform in a wider range around the current anti-node. If there are extra phases arising from inhomogenous coupling between the qubits and TLR, we can correct them with additional single-qubit operations.

Here, we study the one-step generation of high-fidelity GHZ states for many qubits. We did not select a longer resonator with a low frequency of the fundamental cavity mode to solve the problem of the scalability, for the following two reasons. First, the lower frequency of the fundamental cavity mode may lead to more operational errors resulting from the thermal excitations in the cavity. Second, the derived effective Hamiltonian in Eq.~(\ref{effective Hamiltonian}) is valid in the strong-driving regime. If the driving strength in units of Rabi frequency approaches the energy gap of qubits, the counter-wave terms have to be taken into account, as shown in Fig.~\ref{InitialHamiltonian}(b).

To solve the problem of the scalability, let us now use two coupled TLRs as an example to show how the multiparticle GHZ state can be generated via several TLRs in one step. As shown in Fig.~\ref{twocoupledTLR}, $N$ qubits are placed into two cavities formed by two TRLs coupled by a symmetric dc-SQUID. We assume that the qubits labeled from $1$ to $[N/2]$ interacted with TRL A, and other qubits labeled from $[N/2]$ to $N$ are coupled to TRL B, where $[N/2]$ means the maximum integer no more than $N/2$. Near the optimal point and in the basis of the flux qubit persistent current states, the total system Hamiltonian can be given as
\begin{eqnarray}\label{eq:24}
  H_s' &=& \omega_a a^{\dagger}a+\sum\limits_{k=1}^{[\frac{N}{2}]}\frac{\Delta_k}{2}\bar{\sigma}_{x}^k+\sum\limits_{k=1}^{[\frac{N}{2}]}g_k\bar{\sigma}_{z}^k(a^{\dagger}+a)
   \nonumber \\
      & &  +\omega_b
      b^{\dagger}b+\sum\limits_{j=[\frac{N}{2}]+1}^N\frac{\Delta_j}{2}\bar{\sigma}_{x}^j+\sum\limits_{j=[\frac{N}{2}]+1}^Ng_j\bar{\sigma}_{z}^j(b^{\dagger}+b)
      \nonumber \\
      & &+J(\phi_{e})(a^{\dagger}b+ab^{\dagger})+\sum\limits_{k=1}^{N}\Omega_R\cos(\omega_d t)\bar{\sigma}_z^k.
\end{eqnarray}
Here, the operators $a (a^{\dagger})$ and $b (b^{\dagger})$ are annihilation (creation) operators for the field in cavity A with the frequency $\omega_{a}$ and cavity B with the frequency $\omega_{b}$, respectively. $g_{k}$ is the
coupling constant between the $k$-th qubit and the TRL $A$, and $g_{j}$ denotes the coupling of the $j$-th qubit and the TRL $B$.

The parameter $J(\phi_{e})=M_{\rm eff}(\phi_{e})I_{\rm A0}I_{\rm B0}$ is the inductive coupling constant between two TRLs, where $I_{\rm A0}$, $I_{\rm B0}$ are the zero point current in cavity A and cavity B, respectively. It is possible to tune $M_{\rm eff}$ by changing the penetrated flux $\phi_e$ in the dc-SQUID loop, given by~\cite{Brink2005}
\begin{eqnarray}\label{eq:coupler}
  M_{\rm eff}(\phi_{e})&=&-\dfrac{M_{\rm CA}M_{\rm CB}}{L_c}\dfrac{\beta_L\cos(l\pi-\pi\dfrac{\phi_e}{\Phi_0})}{2+\beta_L\cos(l\pi-\pi\dfrac{\phi_e}{\Phi_0})},
\end{eqnarray}
where $l$ is an arbitrary integer, $I_c$ is the critical current of the two identical Josephson junctions in the dc-SQUID, and screening parameter $\beta_L\equiv2\pi L_cI_c/\Phi_0<1$ to ensure that the coupler works in the nonhysteretic regime. We have ignored the direct inductive coupling between two resonators. More detailed discussions of the dc-SQUID coupler can be found in Refs.~\onlinecite{Brink2005} and~\onlinecite{Plourde2004}.

We assume that all qubits are equally driven by the classical field with the frequency $\omega_{d}$, and the coupling between the qubits and the driving field is characterized by the constant $\Omega_{R}$. We rewrite the Hamiltonian in Eq.~(\ref{eq:24}) in the basis of the qubit eigenstates and make the rotating wave approximation, thus the Hamiltonian in Eq.~(\ref{eq:24}) becomes
\begin{eqnarray}\label{eq:25}
   \widetilde{H_1}' &=& \omega_a a^{\dagger}a
   +\sum\limits_{k=1}^{[\frac{N}{2}]}\frac{\Delta_k}{2}\sigma_{z}^k
   +\sum\limits_{k=1}^{[\frac{N}{2}]}g_k(\sigma_{-}^ka^{\dagger}+H.c.)
    \\
      & &  +\omega_b
      b^{\dagger}b+\sum\limits_{j=[\frac{N}{2}]+1}^N\frac{\Delta_j}{2}\sigma_{z}^j
      +\sum\limits_{j=[\frac{N}{2}]+1}^Ng_j(\sigma_{-}^jb^{\dagger}+H.c.)
      \nonumber \\
      & &+J(\phi_{e})(a^{\dagger}b+ab^{\dagger})+\sum\limits_{k=1}^{N}\frac{\Omega_{R}}{2}(\sigma_{+}^{k}e^{-i\omega_d
t} +H.c.).\nonumber
\end{eqnarray}

Without loss of generality, in the following discussion we assume $\omega_a=\omega_b=\omega$, which can be experimentally achieved. Because the TLR is a distributed element, we can easily design its fundamental mode frequency in experiments. Now we introduce a canonical transformation to Eq.~(\ref{eq:25}) via the operators
\begin{eqnarray}\label{eq:26}
  P &=& \frac{1}{\sqrt{2}}(a+b),
  \nonumber \\
  Q &=& \frac{1}{\sqrt{2}}(a-b).
\end{eqnarray}
Since the operators $a$ and $b$ describe different cavity fields in the different cavities, they satisfy the condition $[a,\,b]=0$ and then $[P,\,Q]=0$. Substituting Eq.~(\ref{eq:26}) into Eq.~(\ref{eq:25}) and in the rotating reference frame of the resonant frequency $\omega_d=\Delta$, we have

\begin{eqnarray}\label{eq:27}
   \widetilde{H_2}' &=& (\omega+J)
P^{\dagger}P+\frac{1}{\sqrt{2}}\sum\limits_{k=1}^{[N/2]}g_k(\sigma_{-}^kP^{\dagger}+\sigma_{+}^kP)
   \nonumber \\
      & &  +\frac{1}{\sqrt{2}}\sum\limits_{j=[N/2]+1}^{N}g_j(\sigma_{-}^jP^{\dagger}+\sigma_{+}^jP)
      \nonumber \\
      & &  +(\omega-J)
      Q^{\dagger}Q+\frac{1}{\sqrt{2}}\sum\limits_{k=1}^{[N/2]}g_k(\sigma_{-}^kQ^{\dagger}+\sigma_{+}^kQ)
      \nonumber \\
      & & -\frac{1}{\sqrt{2}}\sum\limits_{j=[N/2]+1}^{N}g_j(\sigma_{-}^jQ^{\dagger}+\sigma_{+}^jQ)
      +\sum\limits_{k=1}^{N}\frac{\Omega_{R}}{2}\sigma_{x}^{k}. \nonumber \\
\end{eqnarray}

Using the same method as for the derivation of the effective Hamiltonian in Eq.~(\ref{effective Hamiltonian}) for the flux qubits coupled to a TLR, based on Eq.~(\ref{eq:27}), we can derive an effective Hamiltonian for the qubits coupled to two TRLs as

\begin{eqnarray}\label{effectiveHcoupledTLR}
H_{\rm eff}' &=&
\sum\limits_{k=1}^{[N/2]}\frac{\sqrt{2}}{4}g_{k}\sigma_{x}^{k}
\left({Pe^{-i(\delta'+J)t}+P^{\dagger}e^{i(\delta'+J)t}}\right)
 \nonumber \\
        & & +\sum\limits_{j=[N/2]+1}^{N}\frac{\sqrt{2}}{4}g_{j}\sigma_{x}^{j}
        \left({Pe^{-i(\delta'+J)t}+P^{\dagger}e^{i(\delta'+J)t}}\right)
        \nonumber \\
        & & +\sum\limits_{k=1}^{[N/2]}\frac{\sqrt{2}}{4}g_{k}\sigma_{x}^{k}
        \left({Qe^{-i(\delta'-J)t}+Q^{\dagger}e^{i(\delta'-J)t}}\right)
        \nonumber \\
        & &
        -\sum\limits_{j=[N/2]+1}^{N}\frac{\sqrt{2}}{4}g_{j}\sigma_{x}^{j}
        \left({Qe^{-i(\delta'-J)t}+Q^{\dagger} e^{i(\delta'-J)t}}\right), \nonumber \\
\end{eqnarray}
with $\delta'=\omega-\omega_{d}$. Here, we already neglect fast oscillating terms under the strong driving condition $\Omega_R\gg g_{k}, \,\delta', J$.

When $\delta'> J,\, g_{k}$, and $\delta'=\xi J\,$( $\xi$ is an arbitrary
odd integer), at the time $t=T_{n}=2\pi n/J$, the
flux qubits are decoupled from resonators. In this case, we have
\begin{eqnarray}
    \sum\limits_{k\neq
    j}^{N}\gamma_{kj}(T_n)
                     &=& \frac{1}{4}\left[\sum\limits_{k,j=1;k\neq
    j}^{[N/2]}\frac{g_{k}g_{j}}{(\delta'+J)(\delta'-J)}\delta' \right.\nonumber \\
    & &+\sum\limits_{k,j=[\frac{N}{2}+1];k\neq
    j}^{N}\frac{g_{k}g_{j}}{(\delta'+J)(\delta'-J)}\delta'
    \nonumber \\
    & &\left.-\sum\limits_{k=1}^{[N/2]}\sum\limits_{j=[\frac{N}{2}]+1}^{N}\frac{g_kg_j}
    {(\delta'+J)(\delta'-J)}J\right]\cdot
    T_n. \nonumber \\
\end{eqnarray}

In the above equation, the first and second terms show that the flux qubits coupled to the same resonator are in permutation symmetry. The third term shows that the coupling between any two qubits mediated by the two coupled TLRs is weaker than the first two terms, and there is also a sign difference between the first two terms and the third term. If we adjust the detuning $\delta'$ and the coupling constant $J(\phi_{e})$ such that the conditions
\begin{equation}
\frac{g_{k}g_{j}}{(\delta'+J)(\delta'-J)}\delta'\cdot
T_n=\frac{1}{2}(3+4m)\pi \label{atsameTLR}
\end{equation}
and
\begin{equation}
\frac{g_{k}g_{j}}{(\delta'+J)(\delta'-J)}J\cdot
T_n=\frac{1}{2}(1+4l)\pi \label{differentTLR}
\end{equation}
are satisfied simultaneously, where $l$ and $m$ are arbitrary integers, then the multiple particle GHZ state can be generated in one step. If we chose $n=1$,  $m=l=0$, $J=2\pi\times40$ MHz, $g_k=\sqrt{2}J$\,($k=1\cdots N$), $\delta'=-3J=-2\pi\times 120$ MHz, the GHZ state can generated in $25$ ns. We would like to emphasize that
$J=2\pi\times40\,$MHz is a reasonable value in experiments. If we assume $I_{\rm Ar0}=I_{\rm Br0}=50$ nA, $J=2\pi\times40\,$MHz means that the effective mutual inductance $M_{\rm eff}$ between two TLRs mediated by dc-SQUID is around $5.32$ pH. It is experimentally realizable with the selected parameters as $I_c=1.5\,\mu$A, $M_{\rm CA}=M_{\rm CB}=60\,$pH and $L_{c}=200\,$pH for the dc-SQUID coupler. It is possible to continuously tune the coupling from anti-ferromagnetic to ferromagnetic like a rf-SQUID coupler in experiments~\cite{Allman2010}. We notice that there is a theoretical work on a flux qubit mediating the coupling between two TLRs and working as a quantum switch~\cite{Mariantoni2008}. In that scheme, it is also possible to realize tunable coupling between two TLRs mediated by the flux qubit.

\begin{figure}
\includegraphics[scale=0.5]{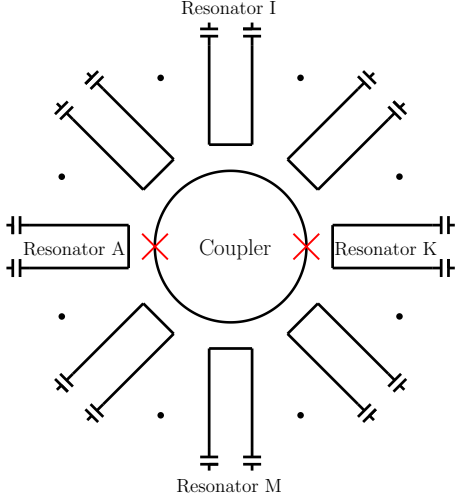}
 \caption{Schematic description of a circle array of TLRs coupled by a dc-SQUID.
 $M$ TLRs are placed as the current anti-nodes around the coupler.
 Flux qubits are coupled to the current anti-nodes of the TLRs. }\label{ScalableTLR}
 \end{figure}

Now, we consider how to generate $N$-qubit ($N>20$) GHZ states in one step. As shown in Fig.~\ref{ScalableTLR}, a dc-SQUID coupler is coupled to $M$ TLRs which form a circle array. Flux qubits are placed at current anti-nodes of the TLRs. For the convenience of discussions, we assume that the coupling constants between resonators and the center coupler are homogenous. The resonators are in permutation symmetry. If we select parameters to simultaneously meet conditions such as those in Eq.~(\ref{atsameTLR}) and Eq.~(\ref{differentTLR}), we can, in principle, extend our scheme to the case of many qubits.

\subsection{Numerical Simulation}
\label{Simulation2}

\begin{figure}
\includegraphics*[scale=0.4]{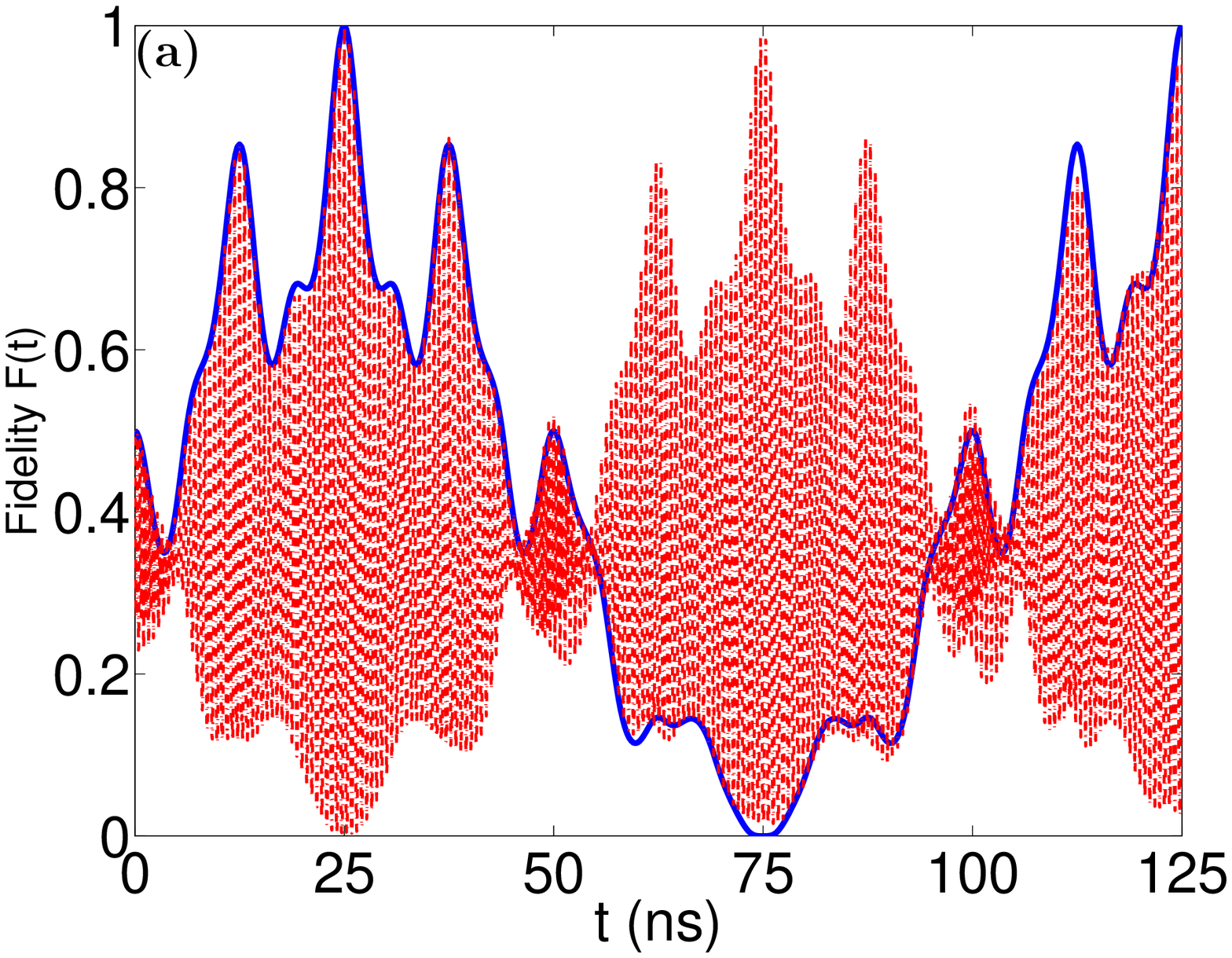}
\includegraphics*[scale=0.4]{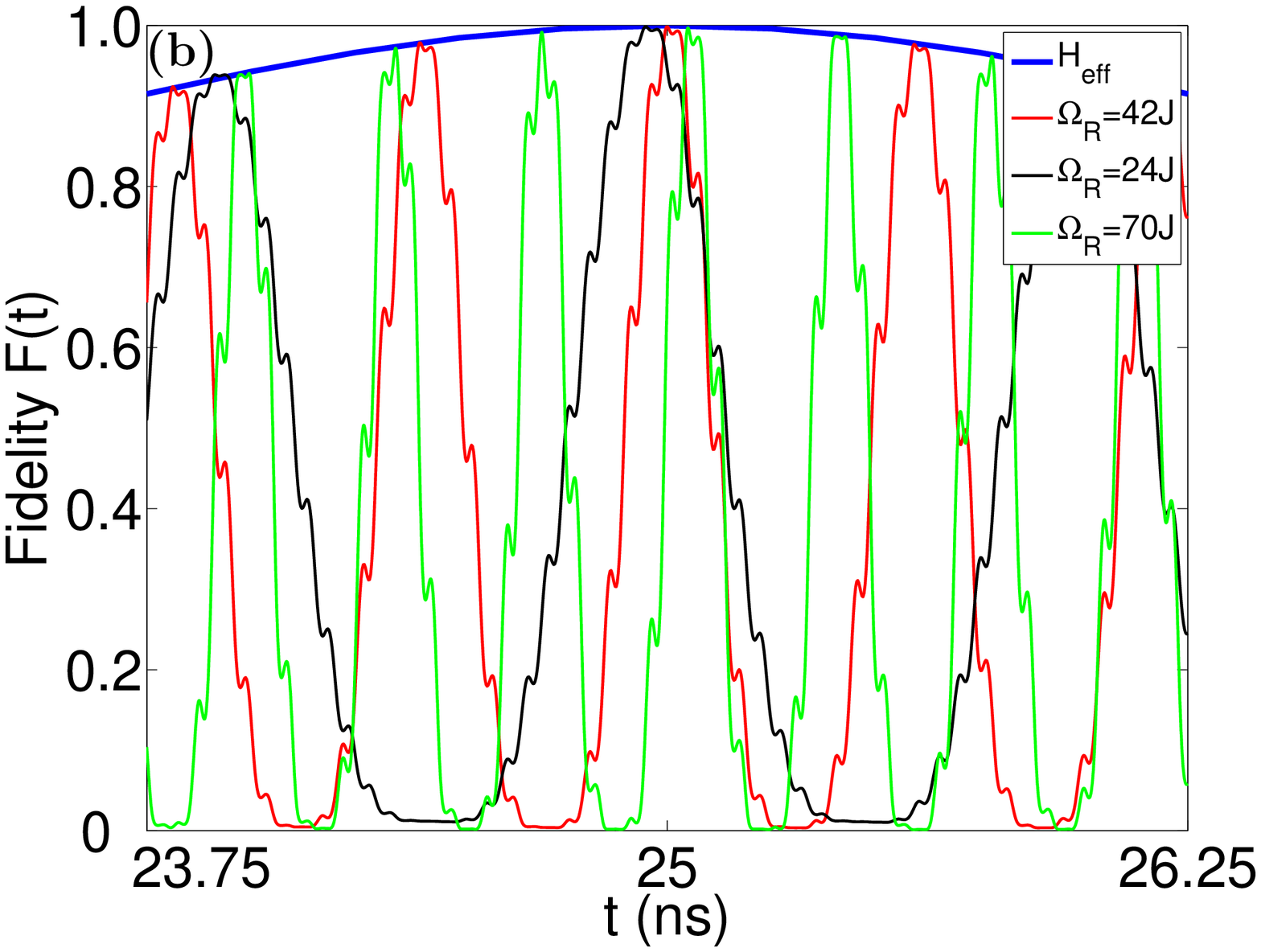}
 \caption{Time dependence of the fidelity of the generated entangled state
for two flux qubits which are coupled by two coupled TLRs. Here, the
two qubits are separately coupled to the two TRLs. We set the following
parameters for our numerical calculations: $\omega_r=2\pi\times10$
GHz, $\Delta=2\pi\times10.12$ GHz, $J=2\pi\times40$ MHz,
$g_{k}=\sqrt{2}J$ and $\delta'=-3J$. The target state is
$(|++\rangle+i|--\rangle)/\sqrt{2}$.
 (a) The blue solid and red dashed curves are simulated using the effective Hamiltonian Eq.~(\ref{effectiveHcoupledTLR}) and the full Hamiltonian in
Eq.~(\ref{coupleHamiltonian}), respectively, with optimized driving strength
 $\Omega_R=42J$. (b) Simulation using the full
 Hamiltonian in Eq.~(\ref{coupleHamiltonian}) with different driving
 strengths around the time $t=2\pi/J=25\,$ns.}\label{SimulationtwoTLR}
 \end{figure}

For flux qubits coupled to a system of two coupled TLRs, the numerical simulation procedure is similar to that in previous case. We can assume that all qubits are initially in the state $\otimes_{k=1}^N|+\rangle_k$ and that the two cavity fields are initially in the vacuum state $|00\rangle_{c}$, i.e., that the initial state of the
whole system is $|\psi(0)\rangle=\otimes_{k=1}^N|+\rangle_k|00\rangle_{c}$. The full Hamiltonian used in the simulation is
\begin{eqnarray}\label{coupleHamiltonian}
   H_{\rm full, sim}' &=&\widetilde{H_2}'+\sum\limits_{k=1}^{N}\frac{\Omega_{R}}{2}
   \left(\sigma_{+}^{k}e^{2i\omega_d t}+\sigma_{-}^{k}e^{-2i\omega_d t}\right)  \\
                  &+&\frac{1}{\sqrt{2}}g_k\sum\limits_{k=1}^{[N/2]}
                  \left(P^+\sigma_{+}^{k}e^{i(\omega_r+\omega_d)t}+H.c.\right)\nonumber \\
                  &+&\frac{1}{\sqrt{2}}g_j\sum\limits_{j=[N/2]+1}^{N}
                  \left(P^+\sigma_{+}^{j}e^{i(\omega_r+\omega_d)t}+H.c.\right)
      \nonumber \\
      &+&\frac{1}{\sqrt{2}}g_k\sum\limits_{k=1}^{[N/2]}
      \left(Q^+\sigma_{+}^{k}e^{i(\omega_r+\omega_d)t}+H.c.\right)\nonumber \\
&-&\frac{1}{\sqrt{2}}g_j\sum\limits_{j=[N/2]+1}^{[N]}
\left(Q^+\sigma_{+}^{j}e^{i(\omega_r+\omega_d)t}+H.c.\right).
\nonumber
\end{eqnarray}

Comparing the simulation results shown in Fig.~\ref{SimulationtwoTLR} with the full Hamiltonian and the
effective Hamiltonian, we find that the effective Hamiltonian in Eq.~(\ref{effectiveHcoupledTLR}) can also describe the dynamics of two qubits system well. With the optimized driving strength, the maximum fidelity can be above $99.8\%$. The two qubits entangled state $(|++\rangle+i|--\rangle)/\sqrt{2}$ can be produced in $t=2\pi/J=25$ ns. Our proposal provides an obvious advantage in that the multiparticle GHZ states are generated in one step in
$\sim 25$ ns with high fidelity. In contrast, in the experiments described in Refs.~\onlinecite{Neeley2010} and~\onlinecite{DiCarlo2010}, three-qubit GHZ state was generated step by step and the total generation time was
$\sim 80$ ns with a fidelity of $\sim 90\%$. According to those experiments, the total generation time will linearly increase with the numbers of qubits.

\section{Discussion and Summary}
\label{Discussion}

Let us now discuss the experimental feasibility. For a TLR with the fundamental mode frequency $\omega_r=2\pi\times10$ GHz and the quality factor $Q\sim10^4$, the decay rate is about $\sim1$ MHz. The
decoherence time achieved in the tunable gap gradiometric qubit is longer than $1\,\mu$s in experiments~\cite{Fedorov2010}. Thus the GHZ state can be generated in about $10$ ns, which is much shorter than the decoherence time of qubits and the decay time in the TLR.

To demonstrate the GHZ state, we could use the well developed quantum state tomography technique in superconducting
qubits~\cite{liu-2005} to reconstruct the density matrix of the final state~\cite{DiCarlo2010,Filipp2009}. After the GHZ state is generated, we can tune the transition frequencies of the qubits such that the interactions between all qubits and the TLRs are switched off. For example, the change of detuning from $2\pi\times 100$ MHz up to $2\pi\times 2$ GHz reduces the coupling strength by a factor of $20$. It is sufficient for experimental demonstration of the GHZ state using quantum state tomography.

We derive the effective Hamiltonian under the strong-driving condition. The strong driving on the flux qubit as high as $2\pi\times2$ GHz has been realized in our group. A detailed theoretical analysis of leakage to higher energy levels under strongly resonant microwave driving is given in Ref.~\onlinecite{Ferron2010}. From these calculation, we can conclude that the leakage to higher energy levels is negligibly small because of the strong anharmonicity for flux qubits at the optimal point ($\varepsilon\approx0$).

In summary, we have proposed a scheme for generation of the multiqubit GHZ entangled state in tunable gap flux qubits coupled to a TLR. We also extend this scheme to the case of tunable flux qubits coupled to two or more coupled TLRs. The operation time is comparable to that of the single-qubit operation. In principle, our
proposal can be generalized to the case in which the tunable qubits are coupled to a circle array formed by $M$ coupled TLRs via a dc-SQUID coupler, that is, our proposal is scalable to some extent. We expect that our scheme has an useful contribution to the generation of cluster states for one-way quantum computation.\\

\appendix

\section{Homogenous coupling between qubits and classical driving field
through TLR} Let us now show how to obtain the Hamiltonian in
Eq.~(\ref{SystemHamiltonian}) when the classical driving field is
applied to the TRL. In our assumption, the driving microwave with large amplitude is substantially detuned from the frequency of resonator. In this situation, quantum fluctuation in the drive is very small compared to the drive amplitude, and the drive field could be considered as a classical field~\cite{Blais2007}. The classical microwave field driving on the resonator can be described by
\begin{equation}\label{drive}
    H_D=\nu(t)a^+e^{-i\omega_dt}+\nu^*(t)ae^{i\omega_dt}
\end{equation}
where $\nu(t)$ is the amplitude and $\omega_d$ is the frequency of external driving. The Hamiltonian of the whole system is $H=H_0+H_I+H_D$, where $H_0$ in Eq.~(\ref{freeHam}) and $H_I$ in Eq.~(\ref{intHam}). We can then displace the field operators using time dependent displacement operator
    \begin{equation}\label{displacement}
        D(\beta)=\exp(\beta a^+-\beta^*a),
    \end{equation}
where $\beta$ is an arbitrary complex number and the field $a$ goes to $a+\beta$ under this unitary transformation.

In the case in which the driving amplitude $\varepsilon$ is independent of time, the displaced Hamiltonian in the energy eigenstate basis of qubits reads
\begin{eqnarray}
    \widetilde{H} &=& D^{+}(\beta)HD(\beta)-iD^+(\beta)\dot{D}(\beta) \\ \nonumber
              &=& \omega_ra^+a+\frac{1}{2}\sum\limits_{k=1}^{N}\Delta_k\sigma_{z}^{k}+\sum\limits_{k=1}^{N}g_k[(a^++\beta^*)+(a+\beta)]\sigma_{x}^{k} \\ \nonumber
              &=& \omega_ra^+a+\frac{1}{2}\sum\limits_{k=1}^{N}\Delta_k\sigma_{z}^{k}+\sum\limits_{k=1}^{N}g_k(a^++a)\sigma_{x}^{k}+\\ \nonumber
              & & \sum\limits_{k=1}^{N}\frac{2g_k\nu\cos(\omega_d t)}{\delta}\sigma_{x}^{k},
\end{eqnarray}
where $\beta(t)=\nu e^{-i\omega_d t}/\delta$, $\delta=\omega_r-\omega_d$.

The above Hamiltonian is the Hamiltonian in Eq.~(\ref{SystemHamiltonian}) in energy eigenstates basis of the qubits with $\Omega_R^k=2g_k\nu\cos(\omega_d t)/\delta$. If the qubits are equally coupled to the resonator, i.e., $g_k\equiv g$, we obtain homogenous coupling $\Omega_R$ between qubits and classical driving field
through the TLR.

\section*{Acknowledgments}
We would like to thank Y.D. Wang, T. Yamamoto,  P.-M. Billangeon, F. Yoshihara, O. Astafiev for their useful discussions. Z.H.P., Y.N. and J.S.T. were supported by NICT Commissioned Research, MEXT kakenhi \textquotedblleft Quantum Cybernetics\textquotedblright and the JSPS through its FIRST Program. Y.X.L. was supported by the National Natural Science Foundation of China under Nos. 10975080, 61025022, and 60836001.

\end{document}